\documentclass[useAMS,usenatbib,referee]{mn2e}
\usepackage{times}
\usepackage{url}
\usepackage{graphics,graphicx}
\usepackage{deluxetable}
\usepackage[usenames]{color}
\DeclareGraphicsExtensions{.pdf,.png,.jpg,.mps,.eps,.ps}
\usepackage{amsmath}
\usepackage{natbib}

\newcommand\asca{{\it ASCA}}
\newcommand\ginga{{\it Ginga}}

\newcommand\chandra{{\it Chandra}}

\newcommand\suzaku{{\it Suzaku}}
\newcommand\rxte{{\it RXTE}}
\newcommand\xmm{{\it XMM-Newton}}

\newcommand\s{{\rm~s}}
\newcommand\ks{{\rm~ks}}

\newcommand\hz{{\rm~Hz}}
\newcommand\mhz{{\rm~mHz}}
\newcommand\kev{{\rm~keV}}

\newcommand\kms{\ifmmode {\rm~km\ s}^{-1} \else ~km s$^{-1}$\fi}
\newcommand\Hunit{\ifmmode {\rm~km\ s}^{-1}\ {\rm Mpc}^{-1}
        \else ~km s$^{-1}$ Mpc$^{-1}$\fi}
\newcommand\ctssec{\ifmmode {\rm~count\ s}^{-1} \else ~count s$^{-1}$\fi}
\newcommand\ergsec{\ifmmode {\rm~erg\ s}^{-1} \else
        ~erg s$^{-1}$\fi}
\newcommand\funit{\ifmmode {\rm~erg\ s}^{-1}\;{\rm cm}^{-2} \else
        ~ergs s$^{-1}$ cm$^{-2}$\fi}
\newcommand\phflux{\ifmmode {\rm~photon\ s}^{-1}\;{\rm cm}^{-2}
        \else   ~photon s$^{-1}$ cm$^{-2}$\fi}
\newcommand\efluxA{\ifmmode {\rm~erg\ s}^{-1}\;{\rm cm}^{-2}\;{\rm
        \AA}^{-1} \else ~erg s$^{-1}$ cm$^{-2}$ \AA$^{-1}$\fi}
\newcommand\efluxHz{\ifmmode {\rm~erg\ s}^{-1}\;{\rm cm}^{-2}\;{\rm
        Hz}^{-1} \else ~erg s$^{-1}$ cm$^{-2}$ Hz$^{-1}$\fi}
\newcommand\cc{\ifmmode {\rm~cm}^{-3} \else cm$^{-3}$\fi}
\newcommand\FWHM{\ifmmode {\rm~FWHM} \else ${\rm~FWHM}$\fi}
\newcommand\Msun{\ifmmode M_{\odot} \else $M_{\odot}$\fi}
\newcommand\Lsun{\ifmmode L_{\odot} \else $L_{\odot}$\fi}
\newcommand\ltsim{\raisebox{-.5ex}{$\;\stackrel{<}{\sim}\;$}}
\newcommand\gtsim{\raisebox{-.5ex}{$\;\stackrel{>}{\sim}\;$}}
\newcommand\hbeta{\ifmmode {\rm H}\beta \else H$\beta$\fi}
\newcommand\Kalpha{\ifmmode {\rm K}\alpha \else K$\alpha$\fi}
\newcommand\nh{\ifmmode N_{\rm H} \else N$_{\rm H}$\fi}
\usepackage{graphicx}
\usepackage{url}
\usepackage{verbatim}
\usepackage{color}

\title[Millihertz QPOs and broad iron line from LMC~X--1]{Millihertz Quasi-Periodic Oscillations and broad iron
  line from LMC~X--1}

\author [Alam et al.] {Md. Shah Alam$^{1}$, G. C. Dewangan$^{2}$,
  T. Belloni$^{3}$, D. Mukherjee$^{2}$, S. Jhingan$^{1}$ \\ 
$^{1}$Centre for Theoretical Physics, Jamia Millia Islamia, New Delhi 110025, India \\
$^{2}$Inter-University Centre for  Astronomy \& Astrophysics (IUCAA), Pune, 411007 India\\
$^{3}$INAF - Osservatorio Astronomico di Brera, Via E. Bianchi 46, I-23807, Merate, Italy; tomaso.belloni@brera.inaf.it}

\begin{document}

\date{Accepted -- . Received -- ; in original form -- }
\pagerange{\pageref{firstpage}--\pageref{lastpage}} \pubyear{????}

\maketitle

\label{firstpage}

\begin{abstract}
  We study the temporal and energy spectral characteristics of the
  persistent black hole X-ray binary LMC~X--1 using two \xmm{} and a
  \suzaku{} observation. We report the discovery of low frequency
  ($\sim 26-29{\rm~mHz}$) quasi-periodic oscillations (QPOs). We also
  report the variablity of the broad iron K$\alpha$ line studied
  earlier with \suzaku{}. The QPOs are found to be weak with
  fractional $rms$ amplitude in the $\sim 1-2\%$ range and quality
  factor $Q\sim 2-10$. They are accompanied by weak red noise or
  zero-centered Lorentzian components with $rms$ variability at the
  $\sim 1-3\%$ level. The energy spectra consists of three varying
  components -- multicolour disk blackbody ($kT_{in} \sim
  0.7-0.9\kev$), high energy power-law tail ($\Gamma \sim 2.4-3.3$)
  and a broad iron line at $6.4-6.9\kev$.  The broad iron line, the
  QPO and the strong power-law component are not always present.  The
  QPOs and the broad iron line appear to be clearly detected in the
  presence of a strong power-law component.
  The broad iron line is found to be weaker when the disk is likely
  truncated and absent when the power-law component almost vanished.
  These results suggest that the QPO and the broad iron line together
  can be used to probe the dynamics of the accretion disk and the
  corona.
\end{abstract}
\section{Introduction}
The highly variable X-ray emission from black hole X-ray binaries
(BHBs) shows a variety of quasi-periodic oscillations (QPOs) that
appear as peaks of small but finite widths in the power density
spectra (PDS)
\citep{Rem99b,2006csxs.book.....L,2006ARA&A..44...49R,2010LNP...794...53B,2011MNRAS.418.2292M}. These
QPOs can be grouped in three categories -- ($i$) high frequency QPOs
that occur in the frequency range of $\sim 30-100\hz$ and are
generally transient \citep[see e.g.,][]{2012MNRAS.426.1701B}, ($ii$)
low frequency QPOs (LFQPOs) in the range of $0.05-30\hz$
\citep{2005ApJ...629..403C,2011MNRAS.418.2292M}, (iii) very low
frequency ($\sim{\rm mHz}$) QPOs that have been observed in the
heart-beat sources GRS~1915+105
\citep{1997ApJ...482..993M,2001MNRAS.322..309T} and IGR~J17091--3624
\citep{2000A&A...355..271B,2011ApJ...742L..17A}.  Recently, mHz QPOs
have also been detected from the BHBs H~1743--322 ($\sim11 \mhz$;
\citealt{2012ApJ...754L..23A}) and IC~X--1 ($\sim 7\mhz$;
\citealt{2013ApJ...771..101P}). Some ultra-luminous X-ray sources
(ULXs) also show very low frequency QPOs at $\ltsim 100\mhz$, e.g.,
M~82~X--1
\citep{2003ApJ...586L..61S,2006ApJ...637L..21D,2006MNRAS.365.1123M,2013MNRAS.436.3262C,2013ApJ...771..101P},
NGC~5408~X--1 \citep{2012ApJ...753..139D}, although they could be the
counterpart of LFQPOs if these ULXs contain intermediate mass black
holes. The LFQPOs come in varieties with three main types -- Type A
(weak with a few percent rms and broad peak around $8\hz$), Type B
(relatively strong with $\sim4\%$ rms and narrow peak around $6\hz$)
and Type C (strong up to $16\%$ rms, narrow and variable peak)
\citep[see e.g.,][]{2005ApJ...629..403C}.

While the exact origin of QPOs from BHBs is still a mystery, the type
C LFQPOs are correlated with energy spectral properties. The centroid
frequency is found to be correlated with the disk flux
\citep{2011MNRAS.418.2292M,1999ApJ...513L..37M}, and the rms amplitude
is found to increase with energy
\citep{2004A&A...426..587C,1997A&A...322..857B,2011BASI...39..409B}.
It implies that the LFQPOs do not directly arise from the thermal
accretion disk as the accretion disk emission does not extend to
higher energies.

The evolution of black hole transients can be characterized in terms
of a limited number of states viz. Low-Hard state (LHS), Hard
Intermediate State (HIMS), Soft Intermediate State (SIMS), High Soft
State (HSS), and the transitions between these states. The LHS is
characterised by a strong dominant power-law emission with a variable
slope ($\Gamma \sim 1.5-2.1$) correlated with flux and a high-energy
cutoff, also variable between 60 and 120 keV \citep[see
e.g.,][]{2009MNRAS.400.1603M}. In the HIMS, a softer component
originating from a thermal accretion disk component contributes more
to the observed emission and the power law component steepens, with an
increase in the high energy cutoff value
\citep{2009MNRAS.400.1603M,2011BASI...39..409B}. The SIMS is softer
than the HIMS due to larger contribution of the disk emission and is
characterised by low level ($\sim$a few $\%$ rms) of variability
\citep{2011BASI...39..409B}.
The LHS is usually associated with a steady radio-jet \citep[][and
references therein]{2013MNRAS.428.2500C}.
In the HIMS, a type-C LFQPO is always observed, as is often the case
in the brightest LHS. In the SIMS, QPOs of type A or B are often
observed. In contrast, LFQPOs are generally not observed in the HSS
dominated by thermal emission from accretion disks \citep[see reviews
by][]{2006ARA&A..44...49R,2010LNP...794...53B}.

Hard X-ray irradiation of the thermal accretion disk can give rise to
X-ray fluorescence emission lines below $10\kev$ and Compton
reflection hump in the $10-50\kev$ band. The iron K$\alpha$ emission
line, broadened by Doppler and gravitational effects near a black hole
in X-ray binaries and active galactic nuclei, is proven to be one of
the most important diagnostic of the inner most regions of strong
gravity \citep[see e.g.,][]{2003PhR...377..389R,2007ARA&A..45..441M}.
The strength and extent of the red wing of the relativistic iron line
is determined by the inner extent of the accretion disk which is
smaller than $6r_g$ if the black hole is spinning, where $r_g=GM/c^2$
is the gravitational radius. For a maximally spinning black hole, the
innermost stable circular orbit is $r_{ISCO} \sim r_g$. Indeed, the
presence of broad iron lines in the X-ray spectra of both black hole
X-ray binaries and AGN have been used to determine the size of the
inner disk and from that to infer the black hole spin.

The production of a broad iron line depends both on the presence of an
accretion disk extending to the innermost regions and a strong hard
X-ray continuum illuminating the disk. The presence of LFQPOs also
depends on the power-law component. This suggests that the iron
K$\alpha$ line and LFQPOs are likely related though not directly.  In
the soft spectral states, strong iron K lines are usually not
observed, due to lack of strong hard X-ray continuum
\citep{2007ARA&A..45..441M}.
Similarly, LFQPOs are not observed in the high/soft states. The
relationship between the broad iron line and the LFQPOs, their
dependence on the hard X-ray continuum and the presence of an
accretion disk are good tests of disk/corona geometry and the models
for the origin of both the iron line and LFQPOs.

LMC~X--1, located in the Large Magellanic Cloud, is a luminous and
persistent black hole X-ray binary (BHB). It consists of a
$10.19\pm1.41{\rm~M_{\odot}}$ black hole primary and an O7~III
companion orbiting each other with a $3.9$-day period
\citep{2009ApJ...697..573O,1995PASP..107..145C}.
The companion drives a strong wind that powers the black hole with an
average luminosity of $0.16L_{Edd}$
\citep{2001MNRAS.320..316N,2009ApJ...701.1076G}.
LMC~X--1 has remained in the HSS persistently and has never been
observed to undergo a transition to the LHS
\citep{2001MNRAS.320..316N,2011ApJ...742...75R}.  The temporal
properties of LMC~X-1 are typical of HSS with its PDS approximately
proportional to $\nu^{-1}$ and fractional root mean square variability
(rms) of $\sim 7\%$ \citep{2001MNRAS.320..316N}. Previously, two QPOs
at $75\mhz$ and $142\mhz$ have been reported from LMC~X--1,
based on {\it Ginga} observations \citep{1989PASJ...41..519E}.
However, a series of nine \rxte{} observations performed in 1996
\citep{1999AJ....117.1292S} and a long $170\ks$ \rxte{} observation
\citep{2001MNRAS.320..316N} did not detect any QPO from the source. It
has been suggested that the $75 \mhz$ and $142\mhz$ QPOs detected by
\citet{1989PASJ...41..519E} is likely an artifact due to incorrect
estimation of the Poisson noise level \citep{2001MNRAS.320..316N}.
The X-ray spectrum of LMC~X--1 is typical of the HSS in which the
thermal disk component dominates over the power-law component
\citep{1989PASJ...41..519E,2001MNRAS.320..316N, 2011ApJ...742...75R}
However, spectral evolution of LMC~X--1 does not follow the modified
Stefan-Boltzmann relation $L_{disk} \propto T_{in}^4$ expected from
the HSS of BHBs \citep[see][]{2011ApJ...742...75R}.
The presence of broad iron line from LMC~X--1, earlier inferred from
\rxte{} observations \citep{2001MNRAS.320..316N}, has been confirmed
by \citet{2012MNRAS.427.2552S} who measured a spin $a
=0.97^{+0.02}_{-0.025}$ using \suzaku{} observations.
\citet{2012MNRAS.427.2552S} also found a strong correlation between
the relative strength of the Compton power law and the iron line flux
using the \rxte{} observations.
 
In this paper, we perform timing and spectral study of LMC~X--1 based
on \xmm{} and \suzaku{} observations and report the detection of
$\mhz$ QPOs and variable broad iron line. We show that the iron line
and the QPO are not always present, and investigate the relationship
between them. We describe the observations analysed and data reduction
in Sect.~\ref{sec:obs_red}, temporal analysis in
Sect.~\ref{sec:timing_ana}, and spectral analysis in
Sect.~\ref{sec:spec_ana}. We finally discuss our results in
Sect.~\ref{sec:discuss}.

\section{Observations \& Data Reduction} \label{sec:obs_red}
\subsection{\xmm{}}
\xmm{} observed LMC~X--1 twice: on 2000 October 21 (obsID: 0112900101;
hereafter {\sc xmm-101}) and 2002 September 26 (obsID:0023940401;
hereafter {\sc xmm-401}) for exposure times of $7.2\ks$ and $40\ks$,
respectively.  The EPIC-pn camera was operated in timing mode using
the thick optical blocking filter in 2000 October and thin filter in
2002 September. The MOS cameras were operated in the full frame mode
using the medium optical blocking filter in both observations. We used
SAS version 12.0 and the most recent calibration database to process
and filter the event data. We corrected the EPIC-pn event list for the
rate-dependent charge transfer inefficiency which has been seen in the
fast mode
data\footnote{\url{http://xmm2.esac.esa.int/external/xmm_sw_cal/calib/index.shtml}}.
We checked for particle background by extracting lightcurves above
$10\kev$ from both EPIC-pn and MOS data. No flaring background was
found in the {\sc xmm-101} data. In the {\sc xmm-401} data, the
flaring particle background was present in the beginning of
observation for a short interval of $\sim1400\s$ and after an elapsed
time of $\sim 26\ks$. We excluded these intervals of high particle
background by using good time intervals created based on count rate
cutoff criteria. We extracted the $10-12\kev$ light curve from the
EPIC-pn data and created a GTI file by selecting the intervals with
the count rate $\le0.5{\rm~counts~s^{-1}}$.
We then applied the GTI file and filtered the event list for high
particle background. This resulted in the net exposures of $23.3\ks$
for the {\sc xmm-401} observation. The MOS data from both observations
were affected with photon pile up. So we did not use MOS data for
further analysis.

In the EPIC-pn timing mode, only one CCD chip is operated. The data
are collapsed into one-dimensional row of size 64 pixels or
$4.4\arcmin$ and are continuously transferred along the second
dimension and readout at high speed resulting in $30\mu\s$ time
resolution. This allows for high count rates and photon pile-up is
negligible below
$800{\rm~counts~s^{-1}}$\footnote{\url{http://xmm.esac.esa.int/external/xmm_user_support/documentation/uhb_2.1/XMM_UHB.html}}. The
EPIC-pn count rate of LMC~X--1 was only
$113{\rm~counts~s^{-1}}$. Hence, the EPIC-pn data in the timing mode
were not affected with pile-up which was also verified with {\tt
  epatplot}.  We extracted the source spectra from the EPIC-pn single
pixel events using a rectangular region of $15$ pixel width covering
the source.  We also extracted the corresponding background spectra
uwing two rectangular regions of widths 10 and 5 pixels away from the
source.  We used the SAS tasks {\tt rmfgen} and {\tt arfgen} to
generate the response files.

\subsection{\suzaku}
\suzaku{} observed LMC~X--1 starting on 2009 July 21, 18:38 UT for
$129.8\ks$ (Observation ID 404061010).  The observation was performed
at XIS nominal pointing and the XIS were operated in the 1/4 window
mode resulting in time resolution of $2\s$. We used \suzaku{} FTOOLS
version 19 and reprocessed and screened the XIS and HXD PIN data using
the {\tt aepipeline} and the most recent version of calibration
database to produce the cleaned event lists. We checked for photon
pile up in the XIS data using the ISIS
tools\footnote{\url{http://space.mit.edu/CXC/software/suzaku/index.html}}
{\tt aeattcor.sl} and {\tt pile\_estimate.sl}. A new attitude file was
created which was then applied to the XIS event lists that resulted in
sharper images. An estimated, minimum pile-up fraction image was
created at different levels $0.03\%$, $0.3\%$, $1\%$, $2\%$, $5\%$,
$10\%$, and $20\%$.  For spectral extraction from the XIS0 data, we
made a circular region file with $130\arcsec$ radius and from its
center excluded a rectangular region ($14.5\arcsec\times32.3\arcsec$)
with pile-up fraction $>5\%$. Similar regions were created for XIS1
and XIS3. The source spectra was extracted from these regions using
xselect. We also extracted background spectra from multiple circular
regions with typical sizes $\sim 60\arcsec$ away from the source. The
response files were created using the tools {\tt xisrmfgen} and {\tt
  xissimarfgen}. The HXD/PIN spectral products were extracted using
the tool {\tt hxdpinxbpi}.

\section{Temporal Analysis} \label{sec:timing_ana}

We used the X-ray timing software
GHATS{\footnote{\url{http://astrosat.iucaa.in/~astrosat/GHATS_Package/Home.html}}}
version 1.1.0 to compute power density spectra (PDS). We begin with
temporal analysis of $0.3-10\kev$ \xmm{} EPIC-pn data. 
The PDS for {\sc xmm-401} were created with a time binsize of
$0.096\s$ (Nyquist frequency of $5.2\hz$), and time segments of 4096
bins in each lightcurve. The PDS of different segments were averaged
and the resulting PDS were logarithmically rebinned in frequency to
improve statistics.  The PDS were computed using rms normalisation
\citep{1990A&A...230..103B}.
The PDS for the {\sc xmm-101} data was derived using a time resolution
of $60{\rm~ms}$ and time segments of 16384 bins in each
lightcurve. The averaged and logarithmically binned PDS for {\sc
  xmm-101} and {\sc xmm-401} are shown in Fig.~\ref{fig:f1}.  The PDS
of LMC~X--1 is featureless red noise during {\sc xmm-101} while a
clear narrow peak in addition to the red noise is seen in the {\sc
  xmm-401} data.

The PDS were fit using ISIS version 1.6.2-27. Unless otherwise
specified, all errors on the best-fit parameters are quoted at the
$90\%$ confidence level corresponding to the minimum
$\chi^2+2.71$. The detection significance of the QPOs were calculated
based on the $1\sigma$ errors corresponding to the minimum
$\chi^2+1.0$. A constant variability power results from Poisson noise
alone, therefore we first fitted a constant model to the PDS derived
from the {\sc xmm-401} data. This model resulted in unsatisfatory fit
($\chi^2/dof=161.5/91$) with large residuals around $\sim$0.03 Hz.
We then added a model for a Lorentzian-shaped QPO for the $\sim$0.03
Hz narrow feature. The parameters of the {\sc QPO} model are the
centroid frequency ($\nu_{qpo}$), the quality factor
($Q=\nu_{qpo}/\Delta\nu$) and the normalization (rms/mean). The {\sc
  constant+qpo} model improved the fit to $\chi^{2}/dof = 97.4/88$,
and resulted in small residuals at the lowest frequencies below
$0.006\hz$. Addition of a powerlaw component ({\sc plaw}) resulted
only in marginal improvement ($\chi^2/dof=91.2/86$). The QPO is
detected at a very high ($6.8\sigma$) statistical significance level,
computed as the ratio between the best fit normalization and its
$1\sigma$ negative error. The best-fit centroid frequency is
$\nu_{qpo}=2.77_{-0.12}^{+0.13}\times10^{-2}\hz$ and the quality
factor $Q= \nu_{qpo}/\Delta\nu = 3.8_{-1.4}^{+3.2}$.  Without the {\sc
  plaw} component, the QPO is detected at much higher ($15.2\sigma$)
significance level and the QPO parameters remained similar.
For {\sc xmm-101}, we found that the {\sc constant+plaw} model
adequately describes the PDS ($\chi^2/dof = 106.3/106$) without any
requirement for a QPO. We calculated an upper-limit on the $rms$ of a
possible QPO by adding a {\sc QPO} model. We fixed the QPO frequency
and the quality factor to the best-fit values obtained for the {\sc
  xmm-401} data and varied the QPO normalization.  This resulted in
$\chi^2/dof=106.3/105$ and the $90\%$ upper-limit on the fractional
rms is $1.1\%$.

To create the PDS of LMC~X--1 using the \suzaku{} XIS data we
extracted lightcurves with $2\s$ bins from the XIS0, XIS1 and XIS3
cleaned data in the $0.4-9\kev$ band and combined the three lightcurves. 
We used the {\sc xronos} task {\sc powspec} to generate
the PDS from the combined XIS lightcurve. We divided the lightcurve in 131 segments of 1024 bins. We
discarded segments with gaps and calculated the PDS from each segment
without any gap and averaged the power in each frequency bin and
obtained the final PDS.  We fitted the PDS in the
$2.5\times10^{-4}-0.25\hz$ range derived from the \suzaku{}/XIS
data. We used a broad Lorentzian, a powerlaw and a constant model for the
continuum as it provided a better fit ($\chi^2/dof = 694.9/506$)
compared to the {\sc plaw + constant} ($\chi^2/dof=725.9/509$). Examination of the
residual showed a narrow peak at $\sim 0.027\hz$ and addition of a
{\sc qpo} improved the fit to $\chi^2/dof=671.8/503$. Thus, we again
detected a QPO at high significance ($6.4\sigma$ level).  The
centroid of the QPO is almost at the same frequency ($\nu_{qpo} \sim
0.027$) as for the {\sc xmm-401} data, and the coherence is high
($Q\sim 4-17$).  We have listed the best-fit PDS parameters for the
three observations in Table~\ref{tab:pds_par}. We have shown the PDS
data and the best-fitting models in the third row of
Fig.~\ref{fig:f1}.  We also created EPIC-pn lightcurve folded
with the corresponding QPO period using the {\sc \sc FTOOLS} task
{\sc efold}.  Fig.~\ref{fig:f2} shows the folded lightcurve.

\begin{table}
  \centering
  \caption{Best-fit model parameter  PDS derived from the \xmm{} observations {\sc xmm-401} and {\sc xmm-101}, and \suzaku{}.} \label{tab:pds_par}
  \begin{tabular}{lllll}
    \hline\hline
    Model & Parameter       &  {\sc xmm-401}          &  {\sc xmm-101}                & \suzaku{}/XIS   \\
    &                       & {\sc const.  + qpo}          & {\sc const + plaw$^a$}                    & {\sc const. + plaw$^a$ + lo + qpo}   \\ 
    \hline
{\sc constant}  &                & $0.0180_{-0.0004}^{+0.0002}$ &     $0.0164\pm0.0003$     & $0.0212_{-0.0007}^{+0.0002}$ \\
{\sc plaw}     & index           & $-0.7_{-3.3}^{+0.4}$          & $-0.85_{-0.16}^{+0.14}$      & $-1.09_{-0.04}^{+0.11}$  \\
               & Norm ($10^{-4}$) & $1.15_{-1.11}^{+5.25}$       & $4.8_{-2.4}^{+3.6}$         &  $2.1_{-0.2}^{+1.5}$     \\ \\

{\sc LO} & $\nu_L$ ($10^{-3}\hz$) &   --                       &      --                 &  $3.1_{-0.7}^{+0.5}$  \\ 
              & FWHM             &   --                       & --                      & $3.1_{-1.7}^{+2.8}$ \\
              & Norm ($10^{-4}$)  &   --                       & --                      &  $4.4_{-1.7}^{+1.7}$  \\ \\
       
    {\sc qpo}  & $\nu_{qpo}$($10^{-2}\hz$) & $2.77_{-0.12}^{+0.13}$  & --               & $2.67_{-0.06}^{+0.07}$\\
               & Q                & $3.8_{-1.4}^{+3.2}$           & --                        & $9.4_{-5.0}^{+8.8}$ \\
               & rms          & $0.019_{-0.003}^{+0.004}$          & --                         & $0.008_{-0.002}^{+0.002}$ \\ \\
    & $\chi^{2}_{min}$/dof      & $91.2/86$                      & $106.3/106$                  & $671.8/503$  \\ 
    & $\Delta\chi^2$ ({\sc QPO})& $-42.0$                    &  --                       & $-23.0$ \\
    \hline
  \end{tabular}
  \\
  $a$ plaw: power-law model \\
  $b$ zfc: Zero-centred Lorentzian model 
\end{table}

\begin{table*}
  % \centering \footnotesize
  \caption{Best-fit spectral parameters of LMC~X--1 derived from the \xmm{} and \suzaku{}  observations} \label{tab:spec_par}
  \begin{tabular}{llccccc} \hline\hline
    Component & Parameter~$^{(a)}$ & \multicolumn{2}{c}{\suzaku{}}      &\multicolumn{2}{c}{\sc xmm-401} & {\sc xmm-101}   \\ 
    &           &    model B~$^{(b)}$  &   model C~$^{(b)}$     & model B~$^{(b)}$ & model C~$^{(b)}$          & model A~$^{(b)}$ \\ \hline
    {\sc tbvarabs} & $N_H$ ($10^{22}{\rm~cm^{-2}}$)    & $1.08_{-0.24}^{+0.24}$ & $1.66_{-0.18}^{+0.15}$   & $1.0_{-0.2}^{+0.3}$ & $1.1_{-0.2}^{+0.4}$    &  $<0.7$  \\ \\
    {\sc diskbb} & $kT_{in}$ (keV)  & $0.79_{-0.01}^{+0.02}$ & $0.77_{-0.01}^{+0.02}$   & $0.65\pm0.02$ & $0.68\pm0.01$       & $0.90\pm0.03$  \\
    & $n_{diskbb}$~$^{(c)}$       & $96.4_{-6.7}^{+12.9}$ & $101.7_{-9.1}^{+18.9}$   & $172_{-51}^{+72}$ & $133.8_{-5.8}^{+4.7}$  & $23.8_{-2.5}^{+5.3}$   \\ \\
    {\sc simpl}  & $\Gamma$ & $2.45_{-0.04}^{+0.04}$  & $2.41_{-0.04}^{+0.06}$   & $2.91_{-0.13}^{+0.04}$ & $2.80_{-0.06}^{+0.04}$  &  $3.3_{-p}^{+p}$ \\
    & $f_{scr}$       &$0.138_{-0.005}^{+0.004}$  & $0.078_{-0.017}^{+0.014}$ & $0.35_{-0.04}^{+0.02}$ & $0.294_{-0.008}^{+0.019}$ & $0.026_{-0.021}^{+0.014}$  \\ \\
    {\sc reflionx} & $n_{ref}~^{(d)}$ ($\times 10^{-7}$)     & -- & $1.17_{-0.16}^{+0.22}$ &--   &  $4.5_{-3.6}^{+p}$ & --\\
    & Fe/solar     & -- & 1 (f)             & -- & 1 (f)            & -- \\
    & $\xi$ ($\rm ergs~cm~s^{-1}$)         & -- & $>8360$               & -- & $1795_{-770}^{+1042}$    & -- \\ \\
         
    {\sc kdblur}  & $i$ & -- & $36.38\degr$ (f)      & --  & $36.38\degr$ (f)    & --\\
    & $r_{in}$ ($r_g$) & -- & $2.38_{-0.19}^{+0.59}$   & -- & $>32$ & -- \\
    & $r_{out}$ ($r_g$) & --  & $400$ (f) & --  & $400$ (f)   &    --  \\
    & $q$            & --  & $3.37_{-0.19}^{+0.25}$   & --  & $3$ (f) & -- \\ \\
    {\sc laor} & $E_{line}$ (keV)  & $6.9_{-0.02}^{+p}$ &  --   & $6.69_{-0.10}^{+0.11}$ & -- & --  \\
    & $r_{in}$   & $2.3_{-0.1}^{+0.1}$ & --   & $64_{-31}^{+113}$ &     -- & -- \\
    & $r_{out}$ ($r_g$) & $400$ (f)  & -- & $400$ (f) &      --   & --  \\
    &  $q$      & $4.5_{-0.1}^{+0.1}$  & --   & $3$ (f)             &  --  & -- \\ 
    & $f_{line}~^{(e)}$ &   $7.9_{-1.0}^{+1.1}\times10^{-4}$  & -- &  $4.6_{-2.0}^{+1.7}\times 10^{-5}$ & -- & -- \\ \\
    &  $\chi^2_{min}/dof$     &  $839.6/731$   & 765.6/731 & 145.5/151 & 145.2/151 & $115.8/107$ \\  \hline

    &  $f_{2.5-10\kev}~^{(f)}$        & $2.4\times 10^{-10}$  & $2.4\times 10^{-10}$    & $2.0\times10^{-10}$ & $2.0\times10^{-10}$ & $1.0\times 10^{-10}$ \\
    &  $f_{10-60\kev}~^{(f)}$         & $7.8\times10^{-11}$   & $7.8\times10^{-11}$     &  --             &  --  &   -- \\
    % $L_X$ & \\
    &   disk frac.~$^{(g)}$          & $\sim 0.60$          & --                   &  $\sim 0.35$      &  --  & $\sim 94$ \\   
    \hline
  \end{tabular}

\noindent
($a$) $p$ indicates that the error calculations pegged at the lower/upper bounds and (f) indicates a fixed parameter. \\
($b$) Models A : {\sc tbvarabs$\times$diskbb$\ast$simpl}; B :  {\sc tbvarabs$\times$(diskbb$\ast$simpl $+$ laor)}; C: {\sc tbvarabs$\times$(diskbb$\ast$simpl $+$ kdblur$\ast$reflionx)}\\
($c$) {\sc diskbb} normalisation $n_{diskbb}=\left(\frac{R_{in}/{\rm km}}{D/10{\rm~kpc}}\right)^2 \cos{i}$, where $R_{in}$ is an apparent inner radius, $D$ is the distance and $i$ is the inclination angle.\\
($d$) {\sc reflionx} normalisation in units of ${\rm photons~cm^{-2}~s^{-1}~keV^{-1}}$ at $1\kev$. \\
($e$) Flux of the {\sc laor} line in units of ${\rm photons~cm^{-2}~s^{-1}}$. \\
($f$) Observed flux in units of ${\rm~ergs~cm^{-2}~s^{-1}}$. \\
($g$) Fraction of disk flux and the {\sc simpl} flux in the $2-20\kev$.
\end{table*}

\section{Spectral Analysis} \label{sec:spec_ana} We used ISIS version
1.6.2-27 for our spectral analysis. As before, we quote the errors on
the best-fit model parameters at the $90\%$ confidence level. In all
spectral fits, we used the absorption model {\sc tbvarabs}
\citep{2000ApJ...542..914W} with fixed abundaces as obtained by
\cite{2010A&A...509L...8H}.

\begin{figure*}
   \centering
   \includegraphics[scale=0.25,angle=-90]{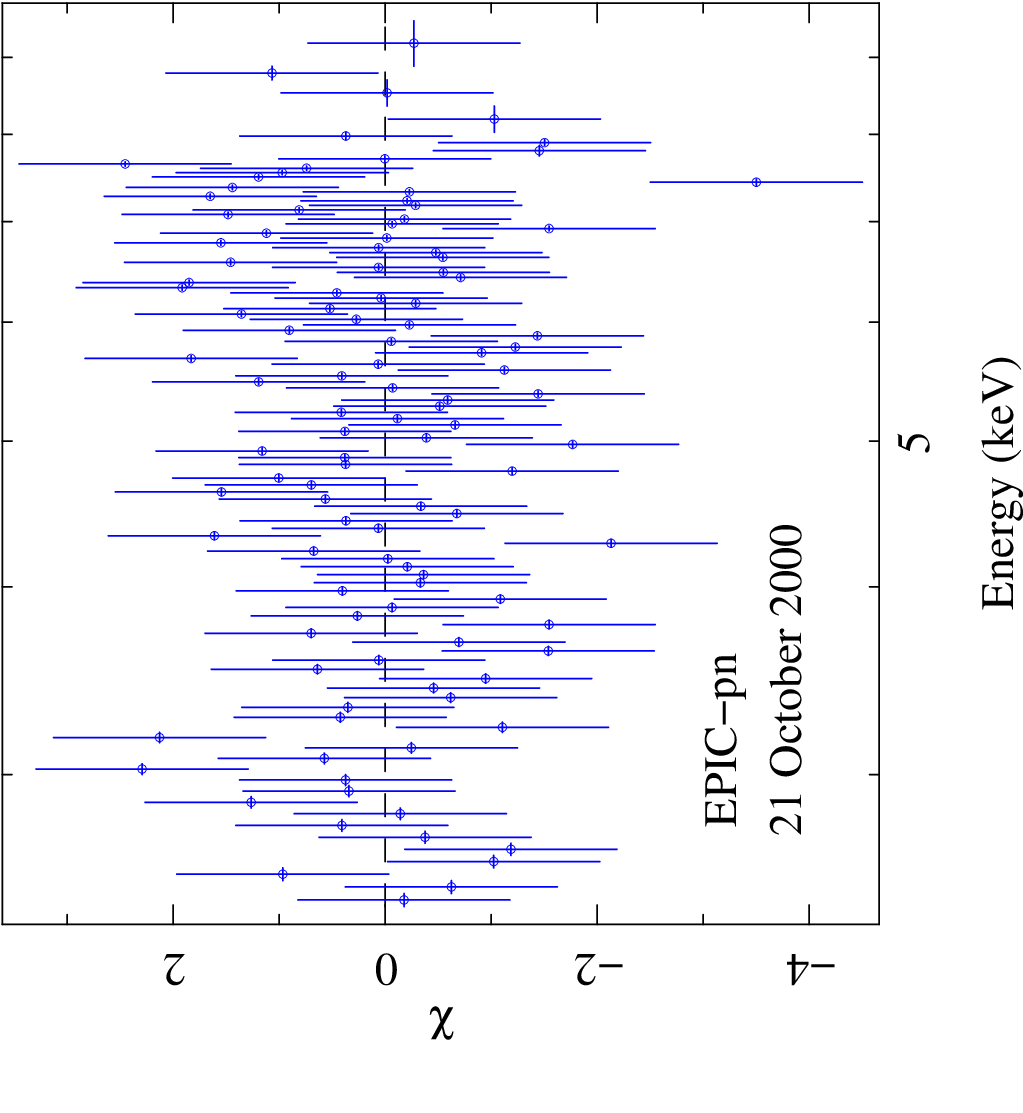}
   \includegraphics[scale=0.25,angle=-90]{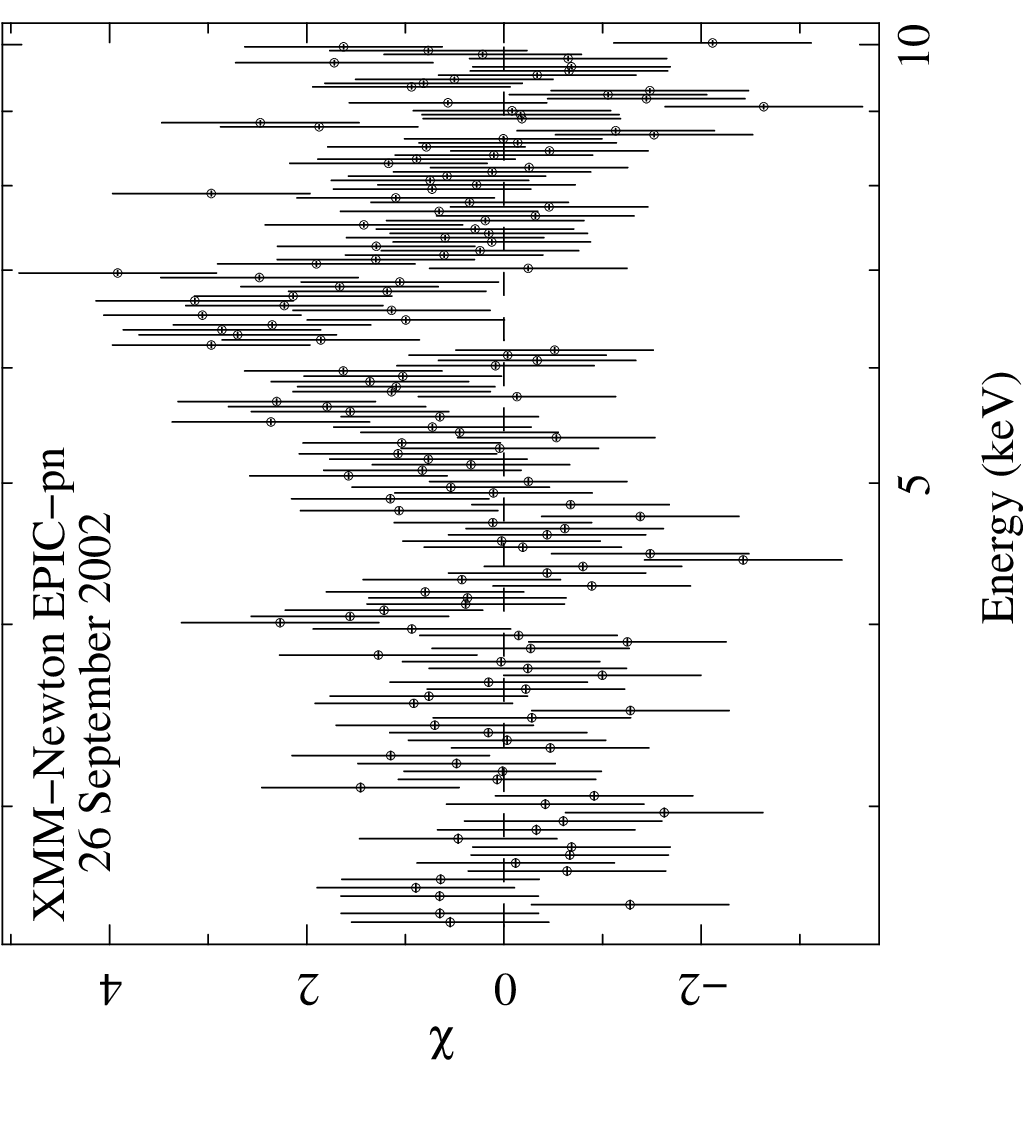}
   \includegraphics[scale=0.25,angle=-90]{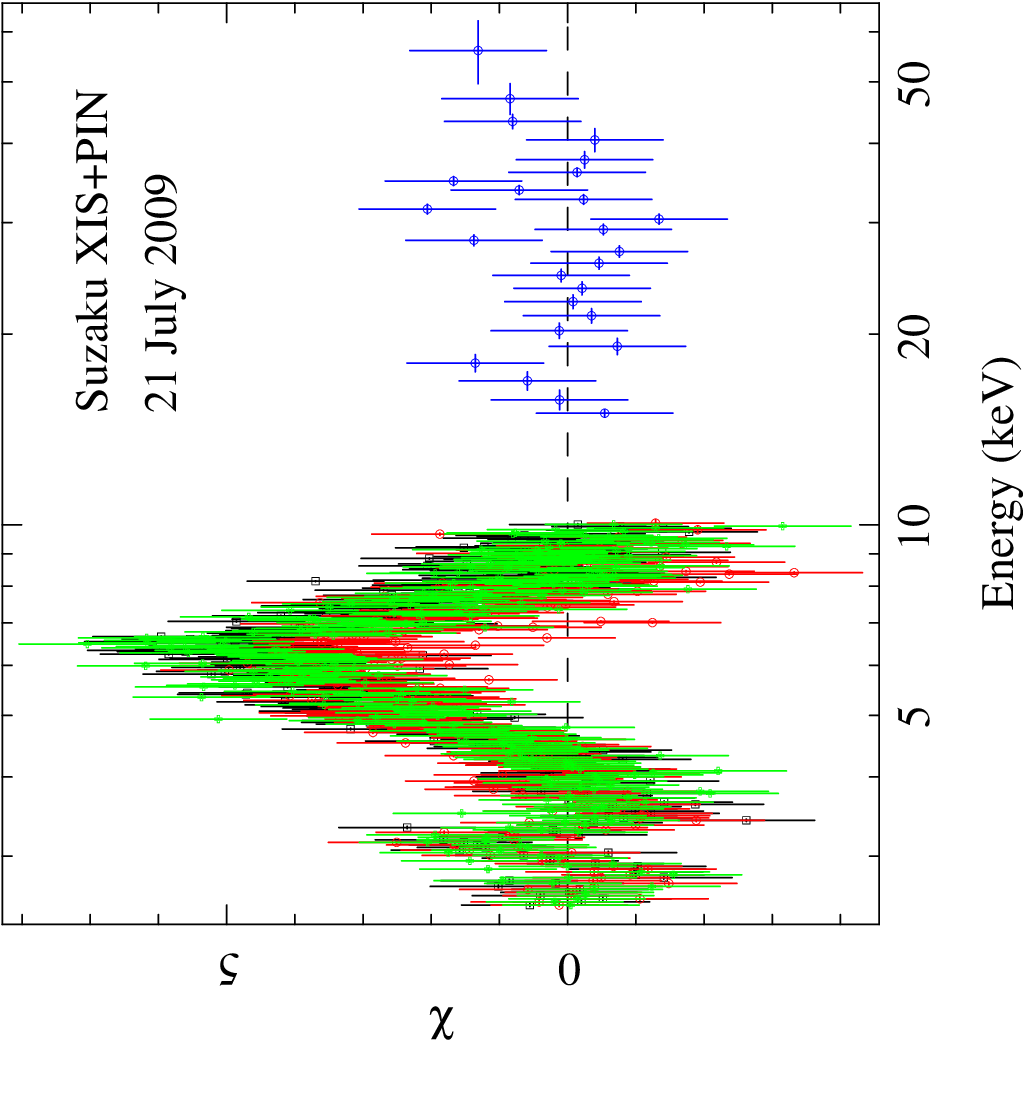}
   \includegraphics[scale=0.25,angle=-90]{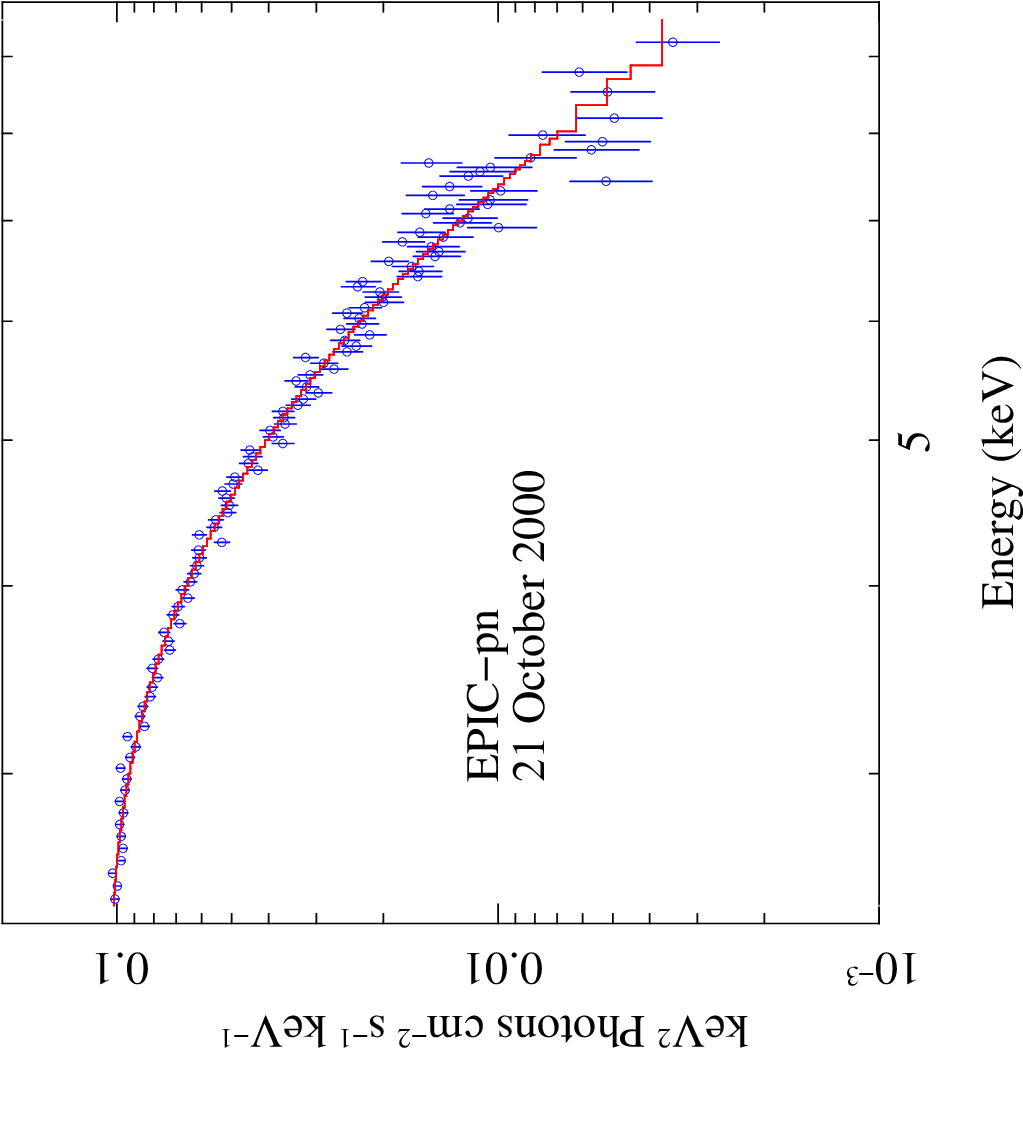}
   \includegraphics[scale=0.25,angle=-90]{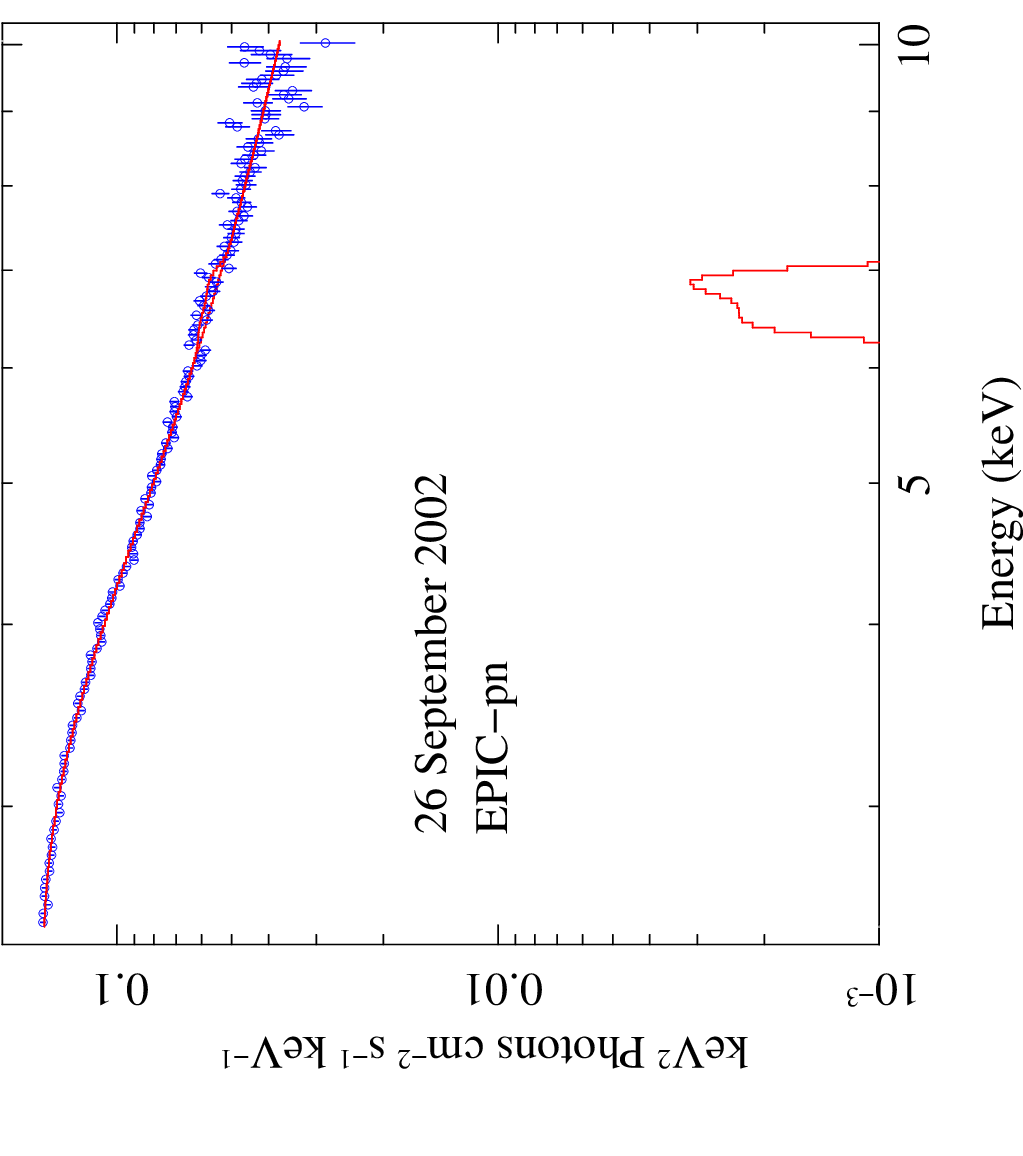}
   \includegraphics[scale=0.25,angle=-90]{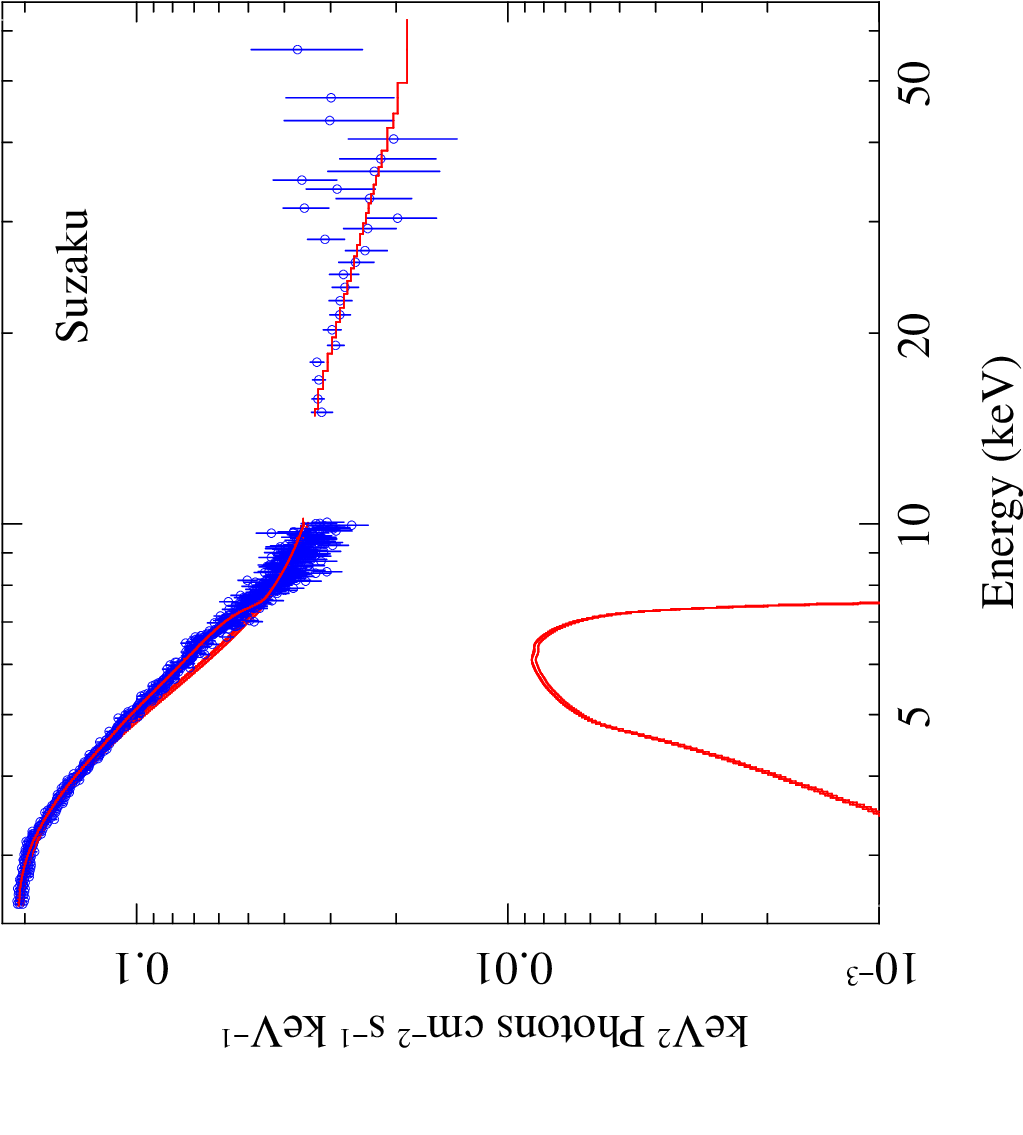}
   \includegraphics[scale=0.25,angle=-90]{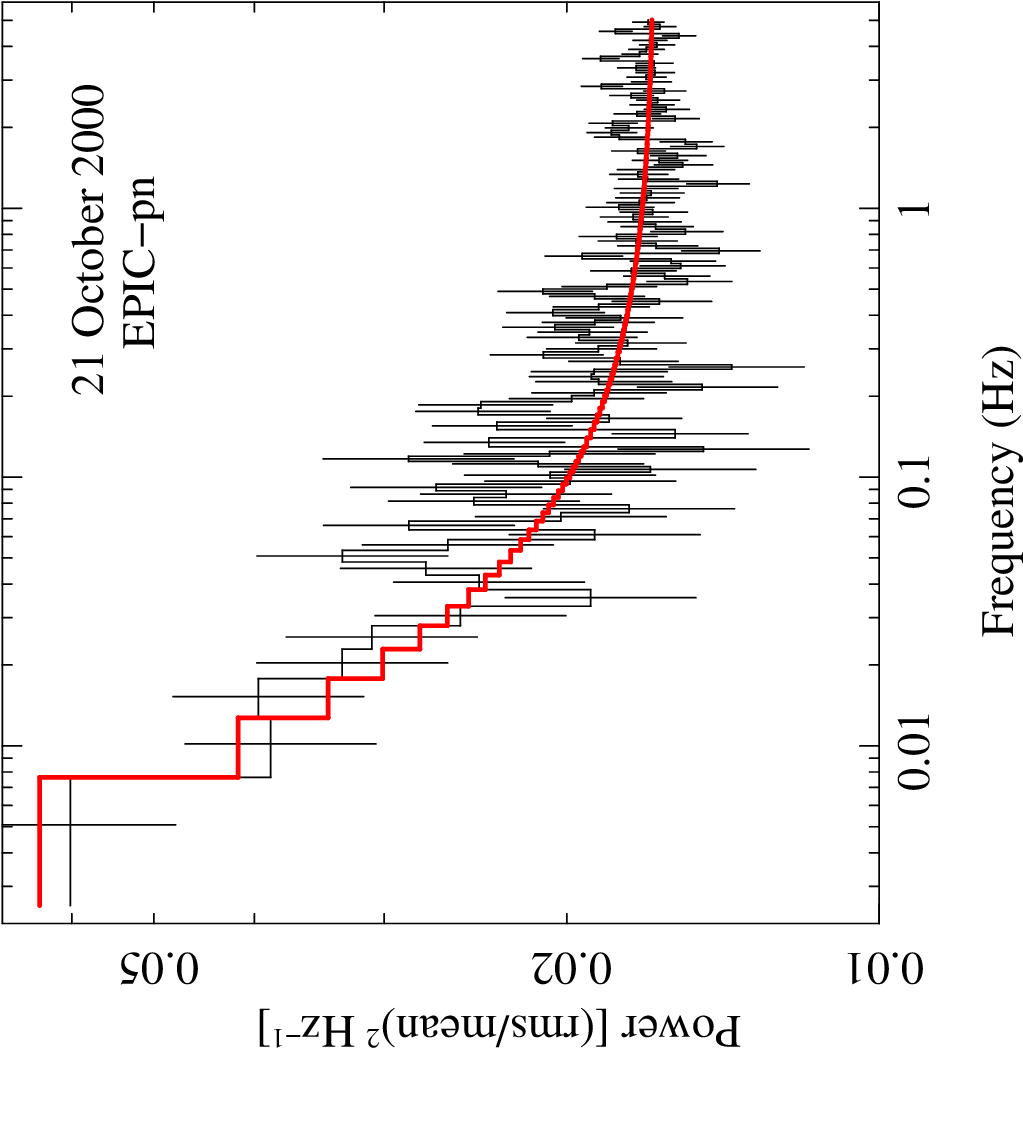}
   \includegraphics[scale=0.25,angle=-90]{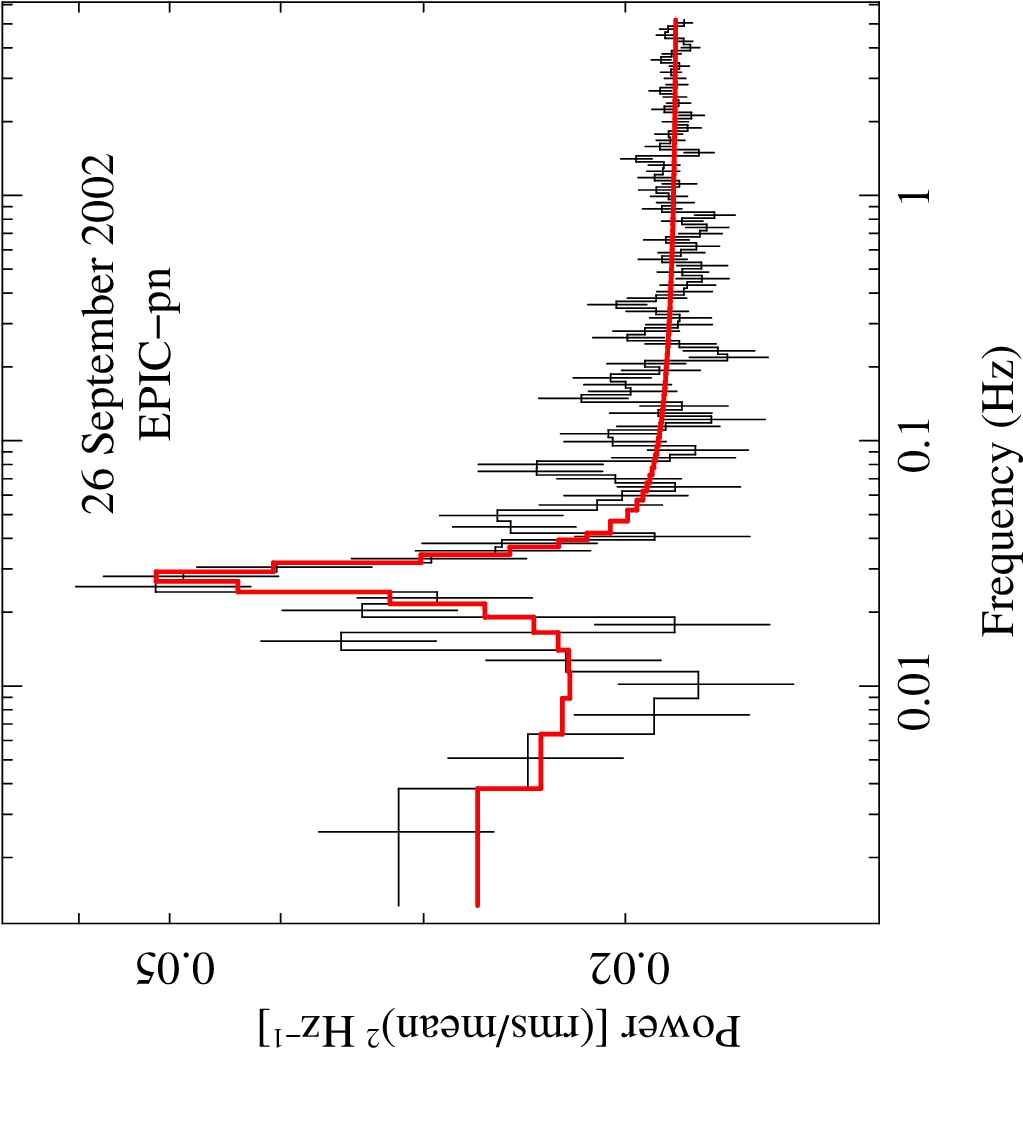}
   \includegraphics[scale=0.25,angle=-90]{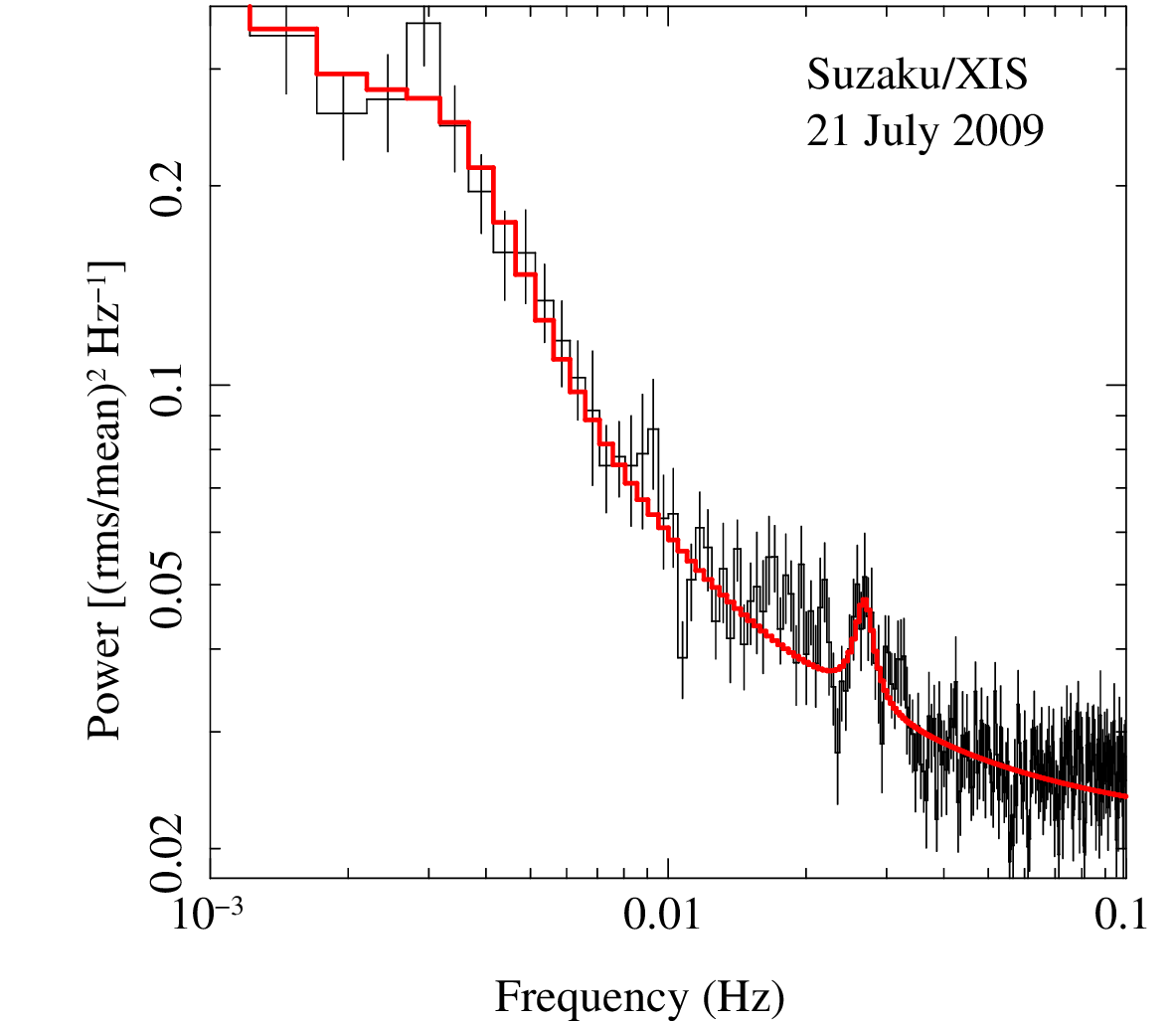}
   \caption{Results of spectral and temporal analysis of \suzaku{} and
     \xmm{} observations of LMC~X--1. {\it Top panels:} Deviations of
     the observed \xmm{} EPIC-pn {\sc xmm-101}, {\sc xmm-401} and
     \suzaku{} XIS+PIN spectral data from the best-fitting {\sc
       tbvarabs$\times$(diskbb + simpl)} models. {\it Middle panels:}
     Unfolded EPIC-pn {\sc xmm-101}, {\sc xmm-101} and \suzaku{}
     XIS+PIN spectral data and the best-fitting models {\sc
       tbvarabs$\times$(diskbb + simpl)} for {\sc xmm-101} and {\sc
       tbvarabs$\times$(diskbb + simpl + laor)} for \suzaku{} and {\sc
       xmm-401} data. {Bottom panels:} The power density spectra and
     the best-fitting models for the {\sc xmm-101}, {\sc xmm-401} and
     \suzaku{}/XIS data.}
   \label{fig:f1}
 \end{figure*}

We began with the spectral fitting of the broadband \suzaku{}
data. The broadband spectrum of LMC~X--1 extracted from the same data
have been studied in detail by \citet{2012MNRAS.427.2552S} who
discovered the relativistic iron K$\alpha$ line and measured the black
hole spin $a_{\star}=0.97_{-0.25}^{+0.02}$. Our purpose here is to
study the spectral variability of LMC~X--1 using \suzaku{} and \xmm{}
observations. We grouped the XIS spectral data to a minimum
signal-to-noise of 10 and minimum channel of 2 in each bin. This
ensures a minimum counts of more than $90$ per bin. To check the
cross-calibration issues between the three XIS instrument we fitted an
absorbed {\sc diskbb} plus {\sc powerlaw} model jointly to the XIS0,
XIS1 and XIS3 data in the $0.5-10\kev$ band.  Examination of the data
to model ratio showed a discrepancy between the three XISs as large as
$20\%$ below $2.5\kev$. Therefore, we excluded XIS data below
$2.5\kev$ and used the XIS data in the $2.5-10\kev$ band. The broad
iron line is clearly seen in the $4-7\kev$ band. There are also slight
cross-calibration problems at a level of $2-3\sigma$ between the back
illuminated CCDs (XIS0 and XIS3) and the front illuminated CCD (XIS1)
in the iron K band. We added a $1\%$ systematic error to each of the
three XIS data sets to account for possible uncertainties in the
calibration of different instruments. We grouped the PIN data to a
minimum signal-to-noise of 3 per bin and used the $15-70\kev$ band. We
fitted the XIS and PIN data jointly with the absorbed {\sc diskbb} and
the empirical convolution model of Comptonization {\sc simpl} which
gives the fraction of photons in the input seed spectrum that are
upscattered into a power-law component \citep{2009PASP..121.1279S}.
Since {\sc simpl} redistributes seed photons to higher energies, we
extended the sampled energies to $1000\kev$ to calculate the model. We
multiplied the model with a {\sc constant} component to account for
variations in the relative normalization of the instruments. The
constant was fixed at 1 for XIS0 and 1.16 for PIN, and varied for XIS1
and XIS3. This model (model A : {\sc
  tbvarabs$\times$(diskbb$\ast$simpl)}) resulted in $\chi^2=1768.9$
for $736$ degrees of freedom (dof). The fit is poor due to the
presence of a broad feature in the $4.5-7\kev$ band reminiscent of
relativistic iron K$\alpha$ line. To show the broad iron line clearly,
we excluded the data in the $4.5-8\kev$ band and refitted. This
resulted in $\chi^2/dof = 355.2/372$ with $N_H\sim
9.6\times10^{21}{\rm~cm^{-2}}$, $\Gamma \sim 2.36$ and $f_{scr} \sim
0.12$. We then evaluated the best-fit model in the $4.5-8\kev$ band.
The upper-right panel in Fig.~\ref{fig:f1} shows the deviations of the
observed data from the continuum model.  To model the relativistic
iron line, we added a {\sc laor} line (model B: {\sc
  tbvarabs$\times$(diskbb$\ast$simpl + laor)}). The {\sc laor} model
describes the line profile from an accretion disk around a spinning
black hole. We fixed the disk inclination at $36.38{\rm~degrees}$ as
determined by \citet{ 2009ApJ...697..573O} based on optical and near
IR observations. We also fixed the outer radius to $r_{out}=400r_g$
and constrained the line energy to vary between $6.4$ and
$6.9\kev$. The fit resulted in $\chi^2/dof=839.6/731$ with
$r_{in}\sim2.3r_g$ and the line energy pegged to the highest allowed
value possibly due to the presence of iron K$\beta$ line.  Since the
broad iron line is thought to be the result of the blurred reflection
from a disk, we replaced the {\sc laor} model with {\sc reflionx}
model which describes the reflection from a partially ionized
accretion disk with constant density as a result of the illumination
of X-ray power-law emission from the corona
\citep{2005MNRAS.358..211R,1999MNRAS.306..461R}. We fixed the iron
abundance at solar, and tied the photon indices of the illuminating
power-law and the {\sc simpl} Comptonization model. We then blurred
the disk reflection with the convolution model {\sc kdblur} to account
for the Doppler and gravitational red shifts. As before, we fixed the
inclination to $i=36.38\degr$ and outer radius to
$r_{out}=400r_g$. This model (model C: {\sc
  tbvarabs$\times$(diskbb$\ast$simpl + kdblur$\ast$reflionx)})
provided a good fit with $\chi^2/dof = 765.6/731$. We have listed the
best-fit parameters for both the models B and C in
Table~\ref{tab:spec_par}. Thus, we confirm the broad iron line earlier
detected by \citet{2012MNRAS.427.2552S} and obtained broadly similar
 iron line parameters though they used a more physical
accretion disk and blurred reflection models and also accounted for
reflection from the stellar wind by using a an ionised reflection
model.

We have also analysed the timing mode EPIC-pn spectral data extracted
from the two \xmm{} observations.
We grouped the EPIC-pn spectral data using the SAS task {\sc
specgroup} to ensure a minimum of 20 counts per bin and at most 5
bins in an FWHM resolution.
First we used the {\sc diskbb} model modified by neutral absorption
({\sc tbvarabs}). As before, we fixed the abundances as obtained by
\citet{2010A&A...509L...8H}. The absorbed {\sc diskbb} model fitted to
the {\sc xmm-101} EPIC-pn spectral data resulted in $\chi^2/dof =
125/109$ with $N_H< 1.5\times 10^{21}{\rm~cm^{-2}}$,
$kT_{in}=0.928\pm0.007\kev$.  The absorption column inferred from the
{\sc tbvarabs$\times$diskbb} model was at least a factor of five lower
than that derived for the \suzaku{} data.  Including the {\sc simpl}
component (model A) resulted in $N_H<0.7\times 10^{22}{\rm~cm^{-2}}$
and improved the fit marginally to $\chi^2/dof=115.8/107$ which
corresponds to $98.3\%$ confidence level according to an F-test. The
best-fit scattering fraction, $f_{pl}=0.026_{-0.021}^{+0.014}$,
indicates that the Comptonising component was extremely weak in the
{\sc xmm-101} observation.

 For {\sc xmm-401}, the absorbed {\sc diskbb} model resulted in a
 statistically unacceptable fit ($\chi^2/dof = 5629.6/156$) due to the
 presence of a hard component. Use of the Comptonization model {\sc
   simpl} (model A) improved the fit ($\chi^2/dof =
 169.4/154$). Careful examination of the fit-residuals showed a hint
 of broad iron line. We excluded the $5-8\kev$ band, performed the
 fitting and compared the observed data with the continuum model. We
 added the {\sc laor} line with the emissivity index fixed at
 $q=3$. The model B improved the fit ($\Delta\chi^2 = -23.9$ for three
 parameters). We also calculated $1\sigma$ error on the line flux and
 found that the broad iron line is detected at a $3.3\sigma$
 level. Next we replaced the {\sc laor} line with {\sc reflionx}
 convolved with {\sc kdblur} (model C) with emissivity index fixed at
 $q=3$. This model also resulted in a good fit
 ($\chi^2/dof=142.2/151$). The best-fit parameters are listed in
 Table~\ref{tab:spec_par}.  For the \suzaku{} and {\sc xmm-401} data,
 the best-fit equivalent Hydrogen column density is in the range of
 $0.8-1.8\times10^{22}{\rm~cm^{-2}}$ while for the {\sc xmm-101} data
 we could only obtain the $90\%$ upper limit ($N_H <
 0.7\times10^{22}{\rm~cm^{-2}}$). These values are generally
 consistent with earlier measurements. Using six soft X-ray spectra
 obtained with grating and/or CCD, \citealt{2010A&A...509L...8H}
 measured $N_H=(1-1.3)\times10^{22}{\rm~cm^{-2}}$ for the {\sc
   diskbb$\ast$simpl} model and $N_H=(1-2)\times10^{22}{\rm~cm^{-2}}$
 for the {\sc diskbb$\times$powerlaw} model, with a systematic
 dependence on the orbital phase. Earlier measurements with
 \chandra{}, {\it BeppoSAX}, \asca{} and {\it BBXRT} span a range of
 $4.4-8.6\times10^{21}{\rm~cm^{-2}}$ (see Table~2 in
 \citealt{2009ApJ...697..573O}).

 \section{Discussion \& Conclusions} \label{sec:discuss}

 We have performed power and energy spectral study of the persistent
 BHB LMC~X--1.  The PDS shape of LMC~X--1 as measured with \xmm{} and
 \suzaku{} is approximately a power-law ($P\propto \nu^{\alpha}$;
 $\alpha \sim -1$), with rms variability of $4.3\%$ ({\sc xmm-101}),
 $2.7\%$ ({\sc xmm-401}) and $4.2\%$ (\suzaku{}) in the $10^{-3}-1\hz$
 range. These values are generally consistent with earlier
 measurements by \ginga{} \citep{1989PASJ...41..519E}, \rxte{}
 \citep[e.g.,][]{2001MNRAS.320..327W} and typical of a HSS.  We have
 discovered QPOs from LMC~X--1 at around $27\mhz$ in \xmm{} ({\sc
   xmm-401}) and \suzaku{} observations. These QPOs appear to be
 remarkable as their centroid frequencies are very low compared to
 that of A, B or C-type QPOs observed from black hole transients in
 outburst. Additionally, they appear in the HSS when QPOs are usually
 not observed.
 The quality factor of these QPOs ($Q\sim 4$ for {\sc xmm-401},
 $\sim10$ for \suzaku{} data) are comparable to that of type B or C
 QPOs and the rms ($\sim 1.1-1.7\%$) are comparable to type A
 QPOs. However, the frequencies are much lower than that observed from
 type A and B QPOs, and the very low frequency type C QPOs are seen
 only in hard states, much different from what we observe here.

 There are very few QPOs from the persistent and wind-fed BHB. In the
 case of LMC~X--1, \citet{1989PASJ...41..519E} claimed detection of
 QPOs at $75\mhz$ and $142\mhz$ with $2.9\%$ and $1.8\%$ rms,
 respectively, based on \ginga{} observations. The weaker QPO is
 likely to be the second harmonic.  Thus, the frequencies of the QPOs
 discovered with \ginga{} are higher but the $rms$ of these QPOs are
 within the range of those measured with \xmm{} and \suzaku{}
 observations. It is likely that all these QPOs are similar in nature
 but vary in their peak frequencies.

 The X-ray energy spectrum of LMC~X--1 is dominated by two spectral
 components -- accretion disk black body and a Comptonising component
 or a steep power-law as was also noted earlier
 \citep[see][]{1989PASJ...41..519E,1994ApJ...422..243S,
   1999AJ....117.1292S, 2001MNRAS.320..316N, 2001MNRAS.320..327W,
   2001MNRAS.325.1253G, 2011ApJ...742...75R, 2012MNRAS.427.2552S}. We
 have confirmed the broad, relativistic iron line from LMC~X--1 in the
 \suzaku{} data, earlier studied in detail by
 \citet{2012MNRAS.427.2552S} who measured the black hole spin
 parameter $a=0.97_{-0.25}^{+0.02}$ ($68\%$ range). We measure the
 inner radius to be $r_{in}=2.4_{-0.2}^{+0.6}r_g$, which corresponds
 to a similar spin parameter though we used a simpler model.  In
 addition, we also detected broad iron line in one of the \xmm{}
 observation ({\sc xmm-401}). However, the line was narrower
 ($r_{in}>32r_g$) and weaker by at least an order of magnitude
 compared to the relativistic line measured with the \suzaku{} data.
 We did not detect an iron line in the {\sc xmm-101} observation. The
 $90\%$ upper limit on the flux of a {\sc laor} line at $6.7\kev$ with
 $r_{in}=64r_g$ (similar to the line in the {\sc xmm-401} data) is
 $3.8\times10^{-5}{\rm~photons~cm^{-2}~s^{-1}}$. This $90\%$ limit on
 the iron line flux is nearly two orders of magnitude lower than the
 line flux measured in the \suzaku{} data but it is formally
 consistent with the $90\%$ range measured with the {\sc xmm-401}
 data.  In any case, the iron line from LMC~X--1 is variable and is
 only detected clearly when a strong power-law component is present.

 We have also found strong variability of the power-law component in
 the energy spectrum. In the first \xmm{} observation ({\sc xmm-101}),
 the power-law component is almost absent. In this observation, the
 fraction of the disk emission in the $2-20\kev$ was $\sim94\%$ and
 LMC~X--1 was in the soft state. In the second \xmm{} ({\sc xmm-401})
 and the \suzaku{} observations, the power-law component was very
 strong with disk fractions $\sim 35\%$ and $\sim 60\%$, respectively,
 in the $2-20\kev$ band (see tab.~\ref{tab:spec_par}).  The spectral
 variability of LMC~X--1 is also well
 studied. \citet{1999AJ....117.1292S} showed that spectral variability
 of LMC~X--1 mainly arises from the changes in the intensity of the
 high-energy power-law component.

 The clear presence of the broad iron line as well as the QPOs in the
 X-ray emission from LMC~X--1 appear to depend on the presence of a
 strong power-law component.
 The QPOs must arise from the inner regions where substantial X-ray
 variability is produced. Indeed, there are theoretical models that
 show oscillations in the inner regions of accretion disks.
 \citet{2000ApJ...542L.111T} have shown that the very low frequency
 QPOs in both BHB and NS XRBs are caused by global disk oscillations
 in the direction normal to the disk. They argue that these disk
 oscillations are the result of gravitational interaction between the
 compact object and the accretion disk. The frequency of the global
 disk oscillations can be written as
 \begin{equation}
   \nu_0 \approx 2\left(\frac{R_{in}}{3R_S}\right)^{-\frac{8}{15}} \left(\frac{M_{BH}}{M_{\odot}}\right)^{-\frac{8}{15}} \left(\frac{P_{orb}}{3hr}\right)^{-\frac{7}{15}}\left(\frac{R_{adj}}{R_{in}}\right)^{-0.3}{\rm Hz}
 \end{equation}
 \citep{2000ApJ...542L.111T}. Using $P_{orb}=3.9{\rm~days}$, $M_{BH} =
 10.9M_{\odot}$ \citep{2009ApJ...697..573O}, $R_{in} = 2.3R_g$ from
 the broad iron line fit to the \suzaku{} data (see
 Table~\ref{tab:spec_par}), and the adjustment radius $R_{adj} =
 2R_{in}$ \citep{2000ApJ...542L.111T}, we find $\nu_0\approx 0.15\hz$.
 Thus, the global disk oscillations are $\sim5$ times faster than the
 observed QPOs. Hence, the global mode oscillations are unlikely to
 explain the observed QPOs from LMC~X--1.

 Very low frequency QPOs in the $\mhz$ range have been observed from
 BHBs e.g., the dynamic QPOs in the $1\mhz-10\hz$ range changing their
 frequency on minutes \citep{1997ApJ...482..993M}, the ``heartbeat''
 QPOs from GRS~1915+105 \citep{2000A&A...355..271B} and
 IGR~J17091--3624 \citep{2011ApJ...742L..17A}. The millihertz
 heartbeat QPOs from GRS~1915+105 and IGR~J17091--3624 occur during
 the high-luminosity, soft-states and are thought to be due to
 limit-cycle oscillations of local accretion rate in the inner disk
 \citep{2011ApJ...737...69N}. The millihertz QPOs from LMC~X--1 are
 also detected in the high luminosity ($\sim 0.5L_{Edd}$) with strong
 soft component. However, the scattering fraction ($f_{SC} \gtsim
 0.13-1$; \citet{2011ApJ...737...69N}) of GRS~1915+105 during the
 heartbeat oscillation appears to be higher than the averaged
 scattering fraction $f_{SC} \sim 0.1$ we found in the presence of QPO
 for LMC~X--1. Also $rms$ amplitude of the QPOs from LMC~X--1 are low
 ($1-2\%$) compared to the $rms$ amplitude of the hearbeat QPOs from
 GRS~1915+105, though the amplitude can be as low as $\sim 3\%$ in the
 case of IGR~J17091--3624 \citep[see
 e.g.,][]{2011ATel.3225....1A}. 
 Moreover, the hearbeat oscillations in
 the $\rho$ and $\mu$ classes of GRS~1915+105 and IGR~J17091--3624
 generally depict a peculiar pattern -- slow rise and fast decay
with  changes in the count rates by factors $\sim 2-5$ \citep[e.g.,][]{2000A&A...355..271B,2011ApJ...742L..17A,2012ApJ...757L..12R}.
 Though it is difficult to infer the pattern of individual variability events, 
 such large amplitude variability are not seen in the lightcurves of LMC~X--1.
 Thus, the $mhz$ QPOs from LMC~X--1 are unlikely to be the
 ``heartbeat'' QPOs.

 Very low frequency, ``non-heartbeat'' oscillations have also been
 found in BHBs. \citet{2012ApJ...754L..23A} have reported $11\mhz$
 QPOs from the Black Hole Candidate H~1743--322 in its successive
 outbursts eight months apart.  These QPOs were almost constant in
 their peak frequency and were detected in the LHS state of
 H~1743--322. Given the differences in the spectral states and
 luminosity of LMC~X--1 and H~1743--322, it is likely that the QPOs
 from these two sources arise from different physical phenomena.

 The QPOs we detected from LMC~X--1 are in the same frequency range as
 those observed in some ULXs, where there are attempts to identify
 them in order to scale with mass. If these are the same features,
 ULXs would host a black hole of a few solar masses.  The $\mhz$ QPOs
 from LMC~X--1 are the only low frequency QPOs from any persistent,
 wind-fed BHB.  A detailed study of the these QPOs and their
 dependence on energy, broadband energy spectral and PDS shape,
 absorption will be required to investigate their nature. The variable
 nature of these QPOs and the broad iron line from LMC~X--1 possibly
 related to strength of the power-law component can be used to study
 the disk-corona coupling and the origin of both these features.  The
 {\it ASTROSAT} mission, scheduled to launch in a year, is well suited
 for detailed study of these QPOs.

 \section*{Acknowledgments}
We thank an anonymous referee for useful comments and suggestions.
 The authors thank M. A. Nowak, R. Misra and A. R. Rao for discussion
 on this work. TMB acknowledges support from grant INAF PRIN
 2012-6. This research has made use of the General High-energy
 Aperiodic Timing Software (GHATS) package developed by T.M. Belloni
 at INAF - Osservatorio Astronomico di Brera. TMB and GCD acknowledge
 support from the joint Indo-Italian project, grant IN12MO12, by the
 Department of Science and Technology (DST), India and the Ministry of
 External Affairs, Italy. This work is based on observations obtained
 with \xmm{}, an ESA science mission with instruments and
 contributions directly funded by ESA Member States and the USA
 (NASA). This research has made use of data obtained from the High
 Energy Astrophysics Science Archive Research Center (HEASARC),
 provided by NASA's Goddard Space Flight Center. SJ and GCD acknowledge support from grant under ISRO-RESPOND program (ISRO/RES/2/384/2014-15).

 \newcommand{\pasp}{PASP} \def\apj{ApJ} \def\mnras{MNRAS}
 \def\aap{A\&A} \def\apjl{ApJ} \def\aj{aj} \def\physrep{PhR}
 \def\pre{PhRvE} \def\apjs{ApJS} \def\pasa{PASA} \def\pasj{PASJ}
 \def\nat{Nat} \def\ssr{SSRv} \def\aapr{AAPR} \def\araa{ARAA}
 \bibliographystyle{mn2e} 
%\bibliography{alam}
 % \bibliographystyle{mn2e} \bibliography{alam}

\label{lastpage}

\end{document}